\documentclass[prl, aps, twocolumn]{revtex4} %preprint,
\usepackage{amssymb}
\usepackage{amsmath,amsthm}
\usepackage{graphicx}
\begin{document}

\title{Heating mechanism affects equipartition in a binary granular system}

\author{Hong-Qiang Wang}
 \email{hqwang@physics.umass.edu}
\author{Narayanan Menon}%
 \email{menon@physics.umass.edu}

\affiliation{%
Department of Physics, University of Massachusetts, Amherst,
Massachusetts 01003-3720
}%

\date{\today}

\begin{abstract}
Two species of particles in a binary granular system typically do
not have the same mean kinetic energy, in contrast to the
equipartition of energy required in equilibrium. We investigate the
role of the heating mechanism in determining the extent of
non-equipartition of kinetic energy. In most experiments, different
species are unequally heated at the boundaries. We show by
event-driven simulations that differential boundary heating affects
non-equipartition even in the bulk of the system. This conclusion is
fortified by studying numerical and solvable stochastic models
without spatial degrees of freedom. In both cases, even in the limit
where heating events are rare compared to collisions, the effect of
the heating mechanism persists.

% 664  chars

\end{abstract}


\maketitle

An issue of broad generality for nonequilibrium physics is the
assignment of intensive temperature variables to steady states of
driven systems. A widely-studied example of practical importance is
a dilute system of macroscopic grains. A granular gas of this kind
is different from an equilibrium gas in that a continuous supply of
energy from an external source is necessary to maintain a steady
state and balance dissipation due to inelastic collisions between
particles. The energy source determines typical scales of the
dynamics in the system, but is presumed not to affect the
microscopic constitutive laws such as the equation of state or
expressions for transport coefficients\cite{BizonGrossman}. In this
Letter we study an externally fluidised granular gas and find that
the details of the energising mechanism affect relations between
intensive quantities even in the bulk of the system.

In a typical real-world situation, energy is delivered at the
boundaries of a granular system by vibration, shear or other
mechanical means. In equilibrium, a gas placed in contact with a
heat bath acquires a uniform temperature and density. However, in
the case of a granular system, there are gradients in density and
particle motions as a function of distance from the energizing
boundary. For dilute gases of inelastic particles there has been
considerable progress \cite{Goldhirsch03}in describing this
inhomogeneous steady state in terms of density and``temperature''
fields, where the temperature is a purely kinetic construct, defined
as the mean kinetic energy per degree of freedom, by analogy to
kinetic theory of a molecular gas.

An important conceptual tool in validating the use of nonequilibrium
temperatures involves thermal contact between systems \cite{Bertin,
Shokef}. One can ask whether a candidate definition for the
temperature governs the direction of energy flow and whether the
temperatures equalise or evolve to bear a fixed relationship. In
this spirit, we consider a granular system with two species of
particles. There is no requirement that the species acquire the same
kinetic temperature; indeed, experiments \cite{Losert99, Klebert,
Wildman} document a violation of equipartition. It is observed that
two species \textit{a} and \textit{b} acquire temperatures whose
ratio $\gamma \equiv T_a/T_b$ is affected strongly by the ratio of
particle masses, $m_a/m_b$, and weakly by their inelasticity. The
temperature ratio $\gamma$ is observed \cite{Klebert} to be only
weakly affected by the composition - the number density and
stoichiometry - of the mixture. These observations have been
reproduced and more comprehensively explored by event-driven
simulations \cite{Barrat02, Paolotti03, Wang} of vibration-fluidised
systems.

This raises the question of whether the extent of departure from
equipartition can be predicted, that is, can the temperature ratio
$\gamma$, be expressed in terms of specified properties of the two
types of particles in the mixture. If so, it would not be necessary
to introduce independent temperature fields to describe each species
in the mixture. Following early work \cite{JenkinsMancini} that
addresses this question within the framework of kinetic theory,
there have been several theoretical treatments of this issue in
spatially homogeneous models. Garz\'{o} and Dufty \cite{Dufty99}
studied theoretically the homogeneous cooling of a binary granular
mixture and found that cooling rates of the two species are the same
while their temperatures remain different, thus maintaining a
time-independent $\gamma$. These predictions have been compared
against MD simulations \cite{Dahl02}.  Likewise, kinetic theory
treatments of homogeneously heated steady-states of mixtures have
been successfully compared to event-driven simulations in which
particles get occasional velocity boosts from an external noise
source \cite{BarratGM02, Pagnani02}. Other treatments of the
homogeneously heated case have been advanced by Langevin-like models
\cite{Morgado} and by mean-field solutions \cite{Marconi02} of the
Boltzmann equation. All of these theoretical results have been
compared to the $\gamma$ found in spatially inhomogeneous
vibration-fluidised steady states \cite{Losert99, Klebert,
Wildman,Barrat02, Paolotti03, Wang}, and while some qualitative
agreement has been found, it is appropriate to be cautious in
comparing these situations.

A further, and often-neglected ingredient in this problem is that in
a typical experiment, the driving mechanism couples differently to
different species of particles.  For instance, in a
vibration-fluidized system \cite{Losert99, Klebert, Wildman}, the
vibrating boundaries may be treated as infinitely massive, and
therefore, impart a characteristic velocity scale (rather than an
energy scale) to the particles. If the particles in the system are
of different mass, then they are differentially heated at the
boundary. This is not an issue in an equilibrium system, where
equipartition holds even if the two species are coupled differently
to a heat bath.  In the granular systems under consideration, the
temperature ratio $\gamma$ is experimentally observed \cite{Klebert}
to be close to the mass ratio of the two species of particles near
the boundary, and then attain an apparently constant value after a
few mean free paths. The question remains whether this apparently
constant bulk value depends on the boundary condition. A similar
issue has previously been commented on in the context of a
homogeneously heated system with both species coupled to different
heat baths\cite{Pagnani02} and in comparing homogeneously heated
systems to boundary-driven systems \cite{Barrat02}.

In this Letter, we address by molecular dynamics simulations the
effect of this differential boundary heating on the extent of
nonequipartition of energy in the bulk of the system for a binary
granular mixture. We find that the level of differential heating
affects nonequipartition in the bulk, and that the effect of the
boundary is never forgotten. These results are strengthened by
consideration of a numerical simulation model \cite{MacKintosh}
without spatial degrees of freedom, and of a stochastic
model\cite{Levine}.

In our event-driven simulations, two types of smooth discs (labelled
$a$ and $b$), both of diameter $d$, but with different masses, are
fluidized inside a rectangle of dimensions $48 \times 32d$. These
simulation parameters are chosen to mimic an experimental geometry
previously studied by us \cite{Klebert}. For the simulations
reported here, the ratio of particle masses $m_b/m_a$ is 5, the
restitution coefficient of both species is 0.93 and an equal number
of both species are used with the total number of particles 2N=200
corresponding to area fraction 10.2\%. We set $g=0$ and employ
periodic boundary conditions in the horizontal ($x$) direction. In
order to isolate the effects we are concerned with, we choose to
heat the particles using fixed``thermal'' walls rather than with the
additional space- and time-dependence introduced by oscillating
walls. Following a collision with the top or bottom boundary, a
particle is reflected with normal and tangential velocity components
independently drawn from gaussian distributions. The width of the
gaussian is chosen to be $<v_x^2>=<v_y^2>=2E_{a,b}/{m_{a,b}}$ for the two types of
particles, thus allowing us to control the differential heating at the boundary.

\begin{figure}
 \centering
 \includegraphics[width=.4\textwidth]{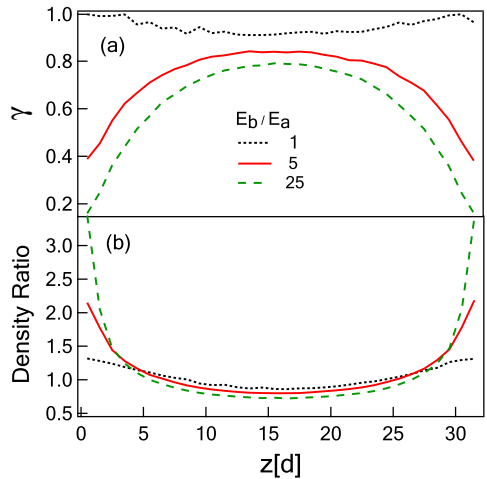}
 \caption{\label{SpacTN}(a)Temperature ratio profile  and (b) number density ratio
 profile for two species of particles along the vertical ($z$) direction. The particles differ
 only in their masses with $m_b = 5m_a$. There are 100 particles
 of each species in a rectangle of size 48 $\times$ 32d, corresponding to an area fraction of 10.2\%.
 Three levels of differential heating are shown:
 $E_a=E_b$(dotted line), $E_a=E_b/5$(solid line), $E_a=E_b/25$(dashed line).}
\end{figure}

In Fig.\ref{SpacTN} we show three choices of differential heating:
$E_b/E_a=1$ where equal amounts of energy are fed to two types of
particles at the boundaries, $E_b/E_a=5$ where the energy fed is
proportional to the mass ratio as in the vibrational experiments,
and $E_b/E_a=25$ which even more disproportionately emphasizes the
heavier particles. Fig.\ref{SpacTN} displays (a) the average
temperature ratio $\gamma=T_a(z)/T_b(z)$  and (b) number density
ratio $n_a(z)/n_b(z)$ profiles along the direction perpendicular to
the top and bottom boundaries. Comparing any single pair of curves -
say the case $E_b/E_a=5$ -  with experimental data \cite{Klebert} we
find similar profiles of $\gamma$ and number density ratio, although
the energy-supplying boundaries are fixed and gaussian. In each of
the three cases, the value of $\gamma$ at the boundaries is
determined by the heating mechanism, as expected, but changes to a
nearly constant value in the bulk.  However, taken together, the
three cases show clearly that the bulk value of $\gamma$ depends on
the choice of differential heating at the boundary. Thus shows that
the value of $\gamma$ is not an intrinsic property of the pair of
particle species and that the effect of the boundary permeates the
bulk of the system.

\begin{figure}
 \centering
 \includegraphics[width=.4\textwidth]{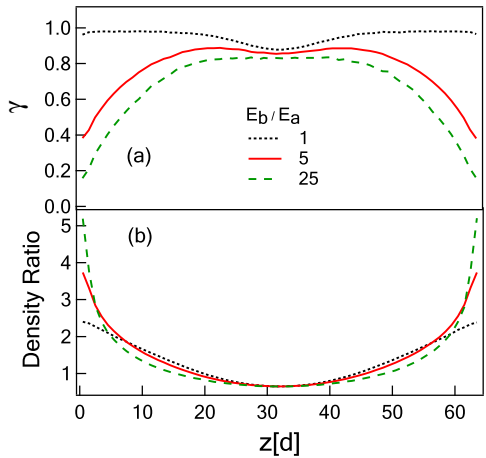}
 \caption{\label{SpacTN64}Vertical temperature ratio profile (a) and density ratio profile (b)
 in a system of height $64d$, double that of the system in Fig.\ref{SpacTN}, with the area fraction
 maintained at 10.2\% The lower panel shows greater segregation of the massive species to the centre
 of the system than in Fig.\ref{SpacTN}.}
\end{figure}

We further check whether the persistence of the boundary condition
into the bulk is a finite size effect by doubling the height of the
system to $64d$ while holding fixed the total area fraction. The
three heating schemes shown in Fig.\ref{SpacTN} are once again
employed, and the temperature ratio and number density ratio as a
function of $z$ are shown in Fig.\ref{SpacTN64}. Clearly, the
temperature ratio $\gamma$ is still affected by the value of
$E_b/E_a$ specified at the boundary.

\begin{figure}
 \centering
 \includegraphics[width=.4\textwidth]{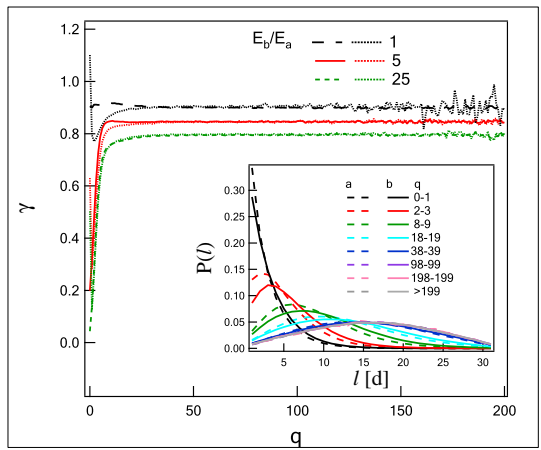}
 \caption{\label{GammaQ} The temperature ratio, $\gamma$, as a function of collision number, q, under
 three levels of differential heating. The corresponding dotted curves sample only those
 trajectories that start at one heating wall and terminate at the other.
 The inset plots the distribution of the distance
 from the last heating boundary \emph{l} for different ranges of $q$.}
% $E_a=E_b$(solid line),  $E_a=E_b/5$(dotted line), $E_a=E_b/25$(dashed line).
\end{figure}

It is straightforward to simulate substantially taller systems,
however, the increasing segregation of the heavier particles to the
middle of the system presents a new complication. We thus address
the issue of whether the data in Fig.\ref{SpacTN} and \ref{SpacTN64}
are still affected by finite-size limitations by considering the
number of collisions $q$ suffered by a particle since it has
collided with a boundary. Each time a particle $i$ collides with
another particle, $q_i$ is incremented by 1, and after it collides
with either heating boundary, $q_i$ is set to zero. The probability
distribution  of the vertical distance $l$ of a particle from the
wall it last collided with, is shown in the inset of
Fig.\ref{GammaQ} for various values of $q$. As expected, P($l$)
broadens and shifts monotonically to larger $l$, for larger $q$,
saturating at the half-height of the system $16d$ for large $q$.
Next, statistics for the kinetic energy of the two types of
particles as a function of $q$ are sampled at fixed time intervals
during a long simulation time and the temperature ratio $\gamma$ is
obtained as a function of $q$. As shown in Fig.\ref{GammaQ}, each
curve of $\gamma (q)$ reaches a plateau after a small number of
collisions. This is consistent with MD simulations of bidisperse
systems\cite{Dahl02} where it was found that after about 10
collisions the temperature ratio $\gamma$ reached a steady-state
value from arbitrary initial conditions. However, as can be seen in
Fig.\ref{GammaQ} the asymptotic value of $\gamma$ at large $q$
remains a function of $E_b/E_a$ . Therefore, details of the driving
mechanism at the boundary appear not to be erased by interparticle
collisions. To ensure that this conclusion is not dominated by
particles that suffered many collisions but happened to remain close
to a boundary, we also plot in Fig.\ref{GammaQ} $\gamma$ for the
subset of particle trajectories that start at one heating wall and
terminate at the other. The plateau values of $\gamma$ of these
trajectories coincide with those from the entire set of particles.

\begin{figure}
 \centering
 \includegraphics[width=.4\textwidth]{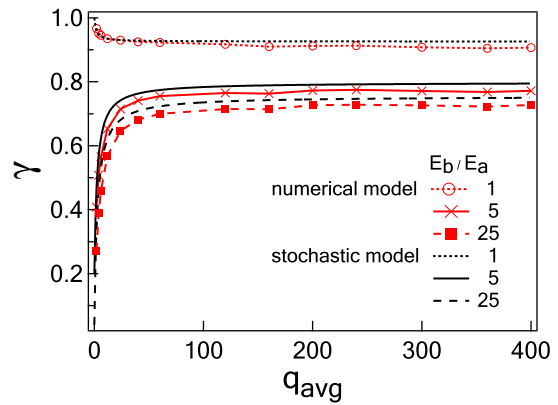}
 \caption{\label{ModelQ} The temperature ratio $\gamma$ as a function of the average particle-particle
 collision times $q_{avg}$. Numerical model results
 are expressed in lines with symbols, $E_a=E_b$(open circle),
 $E_a=E_b/5$(cross), $E_a=E_b/25$(solid square). The smooth curves correspond to results derived from the stochastic model,
 $E_a=E_b$(dotted line),  $E_a=E_b/5$(solid line), $E_a=E_b/25$(dashed line). The corresponding asymptotic values as
 $q_{avg}\rightarrow\infty$ for these
 curves are 0.926, 0.798 and
0.755. }
\end{figure}

It may still be argued that a finite-size effect persists since
particles with large $q$ suffer collisions with particles that have
recently collided with the boundary. In order to further confirm the
persistent effect of the differential heating mechanism we turn to a
numerical model without spatial degrees of freedom, based on that
introduced by van Zon and MacKintosh\cite{MacKintosh} to simulate a
monodisperse system. A system of 2N(N=100) inelastic particles is
initialized with gaussian-distributed initial velocities, and in
each time step, $C$ pairs of particles are randomly selected to
collide in a 2-dimensional inelastic collision, and H particles are
selected for heating. The impact parameters for the collisions are
chosen from a uniform random distribution. For each heating event,
velocities are selected randomly from Gaussian distributions with
average energy $E_a$ and $E_b$, depending on the species of particle
selected. Once again, the three cases of differential heating used
in the spatial event-driven simulations can also be implemented
here. For each of these three cases, statistics for the temperature
ratio $\gamma$ are accrued for many values of $q_{avg}=2C/H$, the
ratio of the frequency of collisions to heating events. (The large
$q_{avg}$ limit corresponds to a boundary driven system in which the
inter-particle collisions are much more frequent than heating
events, and corresponds to large $q$ in Fig.\ref{GammaQ}, without
the complications introduced by spatial gradients). As shown by the
three sets of data symbols in Fig.\ref{ModelQ}, the temperature
ratio $\gamma$ as a function of $q_{avg}$ is different for the three
heating mechanisms even for large $q_{avg}$, where the heating
events are rare relative to the collisions that one might
intuitively expect to erase details of the heating mechanism.

The numerical model discussed above can be described in terms of an
exactly solvable stochastic model \cite{Levine}. We have adapted the
model of Shokef and Levine\cite{Levine} to incorporate equal numbers
of two types of particles $a$ and $b$. The time evolution of the
energy $E^i(t)$ of a particle $i$ of either species during an
infinitesimal time step $dt$ is expressed by a stochastic equation,
\begin{eqnarray}
\label{StoEq}
E^i(t+dt) = \left\{\begin{array}{lll} \underline{\mbox{value:}} &
\underline{\mbox{probability:}}
\\ E^i(t), & 1-\mbox{P}dt,
\\
\\ \frac{\lambda_{ij}(1+\eta)^2}{2(1+\lambda_{ij})^2}E^j(t)-
\\ \frac{2(1+\eta)\lambda_{ij}+1-\eta^2}{2(1+\lambda_{ij})^2}E^i(t)+
\\ (\frac{1+\eta^2}{4}+\frac{1}{2})E^i(t), & (1-f)\mbox{P}dt,
\\
\\E^{i}_a \ or \ E^{i}_b, & f\mbox{P}dt,
\end{array} \right.
\end{eqnarray}
where $E^{i(j)}(t)$ is the instantaneous energy of the particle
$i(j)$ at time $t$, $P$ is the interaction rate per particle per
unit time, $f$ is the fraction selected for external heating,
$\lambda_{ij}$=$m_i/m_j$, $\eta$ is the restitution coefficient,
$E^{i}_{a,b}$ is the external energy given to particle $i$ for a
heating event with an input energy $E^i_{a}$ or $E^i_{b}$ depending
on the species of particle $i$. Therefore, the first line on the RHS
of Eq.\ref{StoEq} corresponds to particle $i$ not undergoing
interaction, the second line corresponds to particle $i$ being
selected for collision with another particle $j$ (averaged over
collisions with species $a$ and $b$) and the third line corresponds
to particle $i$ being selected for heating.  For the second line, we
use a simple one-dimensional collision model with uncorrelated
pre-collisional velocities. In steady state, the ensemble averages
obey $<E^i(t+dt)>=<E^i(t)>$. By solving this stochastic equation, we
determine the dependence of the temperature ratio $\gamma$ on
$q_{avg}$ as shown in Fig.\ref{ModelQ}. To make comparison to the
numerical simulations of the random heating model, we make the
correspondence $f=H/(H+2C)=1/(1+q_{avg})$. Once again, three
different values of $E_b/E_a$ are shown. Despite the simplified
treatment of collisions within our adaptation of the Shokef-Levine
model, we obtain reasonable agreement between the numerical model
and this stochastic model. Qualitatively, the dependence on
$E_b/E_a$ follows the same trend, and the smooth curves in
Fig.\ref{ModelQ} for the three heating conditions do not converge at
large $q_{avg}$.

In summary, we have demonstrated by three independent means, the
persistent effect of the boundary heating mechanism on the extent of
nonequipartition in a binary granular system.  The specific context
for studying the effect of differential heating come from studies of
vibration-fluidized granular systems, however, this type of
differential heating should be generic to other forms of driving: in
shear, for example, particles will be differentially excited
depending on their frictional properties, shape, or size relative to
the features of the shearing surface. Similar effects may be
anticipated even in monodisperse systems where the equipartition
between rotational and translational degrees of freedom may not be
determined purely by particle properties but by the degree to which
each degree of freedom is pumped by the driving mechanism. More
generally, details of the boundary heating mechanism can not be
ignored in describing inelastic gases, leading to concerns about
quantitative comparisons between theories of homogeneously excited
granular systems and boundary-driven experiments.

We acknowledge gratefully useful inputs from  J.L. Machta, K. Facto,
K. Feitosa, Y. Shokef, A. Puglisi and D.C. Candela and financial
support through NASA NNC05AA35A and NSF DMR 0606216.

%====End=======================================================================

\addcontentsline{toc}{chapter}{\rm {\bf BIBLIOGRAPHY}  . . . . . .
. . . . . . . . . . . . . . . . . . . . . . . . . . . . . . . . .}


\begin{thebibliography}{99}

\bibitem{BizonGrossman} C.Bizon, M.D.Shattuck, J.B.Swift, and H.L.Swinney, Phys. Rev. E,
\textbf{60}, 4340(1999); E.L.Grossman, T.Zhou, and E.Ben-Naim, Phys.
Rev. E, \textbf{55}, 4200(1997).


%\bibitem{Grossman} E.L.Grossman, T.Zhou, and E.Ben-Naim,
%Phys. Rev. E, \textbf{55}, 4200(1997).

\bibitem{Goldhirsch03}I.Goldhirsch, Annu. Rev. Fluid Mech.
\textbf{35}, 267(2003).

\bibitem{Bertin} E.Bertin, K.Martens, O.Dauchot, and
M.Droz, Phys. Rev. E \textbf{75}, 031120(2007).

\bibitem{Shokef}Y.Shokef, G.Shulkind and D.Levine, Phys. Rev. E,
\textbf{76},030101(2007).

\bibitem{Losert99} W.Losert \emph{et al}., Chaos \textbf{9}, 682(1999).

\bibitem{Klebert} K.Feitosa and N.Menon, Phys. Rev. Lett.,
\textbf{88}, 198301(2002).

\bibitem{Wildman} R.D.Wildman and D.J.Parker, Phys. Rev. Lett. \textbf{88},
064301(2002).

\bibitem{Barrat02}A.Barrat and E.Trizac, Phys. Rev. E \textbf{66},
051303(2002).

\bibitem{Paolotti03}D.Paolotti, C.Cattuto, U.Marini Bettolo Marconi, A.Puglisi,
Granular Matter \textbf{5}(2), 75(2003).

\bibitem{Wang} H.-Q. Wang, G.-J. Jin, and Y.-Q. Ma,
Phys. Rev. E \textbf{68}, 031301(2003).

%\bibitem{shear} J.M.Montanero and V.Garz\'{o}, Molecular Simulation
%\textbf{29}(6-7), 357(2003).

\bibitem{JenkinsMancini}J.T. Jenkins and F. Mancini, J. Appl. Mech, \textbf{54}, 27
(1987).

\bibitem{Dufty99}V.Garz\'{o} and J.Dufty, Phys. Rev. E \textbf{60},
5706(1999).

\bibitem{Dahl02} S.R.Dahl, C.M.Hrenya, V.Garz\'{o} and J.W.Dufty,
Phys. Rev. E \textbf{66}, 041301(2002).

\bibitem{BarratGM02}A.Barrat, E.Trizac, Granular Matter
\textbf{4}(2), 57(2002).

\bibitem{Pagnani02}R.Pagnani, U.Marini Bettolo Marconi and
A.Puglisi, Phys. Rev. E \textbf{66},051304(2002).

\bibitem{Morgado}W.A.M.Morgado, Physica A \textbf{320},60(2003).

%This is for a steady state with stochastic heating
\bibitem{Marconi02}U.Marini Bettolo Marconi and A.Puglisi, Phys. Rev. E
\textbf{66}, 011301(2002)
%;This is the free cooling paper
%U. Marini Bettolo Marconi and A. Puglisi, Phys. Rev. E\textbf{65},
%051305(2002)


\bibitem{MacKintosh} J.S.van Zon and F.C.MacKintosh,
Phys. Rev. Lett. \textbf{93}, 038001(2004); Phys. Rev. E
\textbf{72}, 051301(2005).

\bibitem{Levine} Y. Shokef, D. Levine, Phys. Rev. E \textbf{74}, 051111(2006); Y.Srebro and D.Levine, Phys. Rev. Lett.,
\textbf{93}, 240601(2004).

%--------------------------------------------------------------








\end{thebibliography}
\end{document}